\documentclass[11pt,twoside]{article}
\usepackage{asp2006}
\usepackage{epsfig}
\usepackage{epsf}
\usepackage{graphics}
\usepackage{lscape}
\def\edcomment#1{\iffalse\marginpar{\raggedright\sl#1\/}\else\relax\fi}
\reversemarginpar

\newcommand{\etal}{et~al.}
\newcommand{\uv}{({\em u,v})}
\newcommand{\water}{H$_2$O}

\newcommand{\dfig}[2]            
{
  \begin{center}
    \begin{minipage}[t]{0.6\textwidth}
        \psfig{file=#1,height=0.5\textwidth}
    \end{minipage}
    \hfill
    \begin{minipage}[t]{0.3\textwidth}
        \psfig{file=#2,height=0.9\textwidth}
    \end{minipage}
  \end{center}
}

\newcommand{\plottwoone}[2]{%
 \centering 
 \leavevmode 
 \columnwidth=.6\columnwidth 
 \includegraphics[width={\eps@scaling\columnwidth}]{#1}%
 \hfil 
 \columnwidth=.5\columnwidth 
 \includegraphics[width={\eps@scaling\columnwidth}]{#2}%
}%

\begin{document}
\title{Conclusions from the image analysis of the VSOP Survey}
\author{R. Dodson}
\affil{Observat\'orio Astronomico Nacional, Madrid,
Spain}
\author{E. Fomalont}
\affil{NRAO, USA}
\author{K. Wiik}
\affil{Tuorla Obs., Finland}
\author{VSOP Survey Working Group}
\affil{}

\begin{abstract}
  In February 1997, the Japanese radio astronomy satellite HALCA was
  launched to provide the space-bourne element for the VLBI Space
  Observatory Programme (VSOP) mission. A significant fraction of the
  mission time was to be dedicated to the VSOP Survey of bright
  compact Active Galactic Nuclei (AGN) at 5~GHz, which was lead by ISAS.
  The VSOP Survey Sources are an unbiased dataset of 294 targets, of
  which 82\% were successfully observed. These are now undergoing
  statistical analysis to tease out the characteristics of typical AGN
  sources. We present here the summary of the imaging and conclusions
  we have reached.

\end{abstract}

\vspace{-0.5cm}
\section{Introduction}

The radio astronomy satellite HALCA (Highly Advanced Laboratory for
Communications and Astronomy) was launched by the Institute of Space
and Astronautical Science in February 1997 to participate in Very Long
Baseline Interferometry (VLBI) observations with arrays of ground
radio telescopes.  HALCA provides the longest baselines of the VSOP,
an international endeavour that has involved over 28 ground radio
telescopes, five tracking stations and three correlators
\citep{hir98,hir00a}.  HALCA was placed in an orbit with an apogee
height above the Earth's surface of 21,400\,km, a perigee height of
560\,km, and an orbital period of 6.3~hours.

During the seven years of HALCA's mission lifetime, most of the
observing time was used for General Observing
Time (GOT). The remaining observing time was devoted to a mission-led
survey of active galactic nuclei at 5\,GHz: the VSOP Survey Program.
The major goal of the Survey was to determine the statistical
properties of the sub-milliarcsecond structure of a complete sample of
AGNs.  \citep{hir00b, fom00a}.  Following the end of the formal
international mission period in February 2002, the Japanese-dominated
effort continued survey observations until October 2003, when HALCA
lost its attitude control capability.

This paper presents the summary of the Survey imaging analysis, which has
been completed, and some early conclusions from the statistical
analysis. Early papers in the VSOP Survey series include \citet{sco04}
and \citet{dod08}, which present the 242 images and models (P-III and P-V), and
\citet{hor04} (P-IV) which presents the statistical conclusions based
on data in P-III. They analyzed the cumulative visibilities of those
sources to obtain the `typical source structure'. We repeated this with
the entire sample of sources, and discuss the the brightness
temperature properties.

\section{The Observations}

The VSOP mission and the 5 GHz AGN Survey are
fully discussed in \citet{hir98,fom00b,hir00a,hir00b}.
Briefly, in order to be included in the VSOP Survey, a
source was required to have:
\newline $\bullet$ a total flux density at 5 GHz, $S_5\geq
5.0$~Jy\newline \centerline{or\phantom{aaaaaaaaaaaaaaaaaaaaaaa}}
$\bullet$ a total flux density at 5 GHz, $S_5\geq 0.95$~Jy and\newline
$\bullet$ a spectral index $\alpha \geq -0.45$ ($S \propto
\nu^\alpha$) and\newline $\bullet$ a galactic latitude $|b|\geq
10^\circ$.

The finding surveys from which sources were selected were primarily
the Green Bank GB6 Catalog for the northern sky \citep{gre96}, and the
Parkes-MIT-NRAO (PMN) Survey \citep{law86,gri93} for the southern sky.
As this was compiled from single dish catalogues, some
of the selected sources would not be detectable by HALCA due to
insufficient correlated flux density on baselines longer than about 1000
km.  Therefore, sources with declination $>-44^\circ$ were observed in
a VLBA pre-launch survey (VLBApls, 
Fomalont et al. 2000b) and a cutoff criterion, a minimum flux density
of $0.32$\,Jy at 140~M$\lambda$, was established for inclusion of a
source in the VSOP Survey \citep{fom00a}. For sources south of
$-44^\circ$ this cutoff could not be determined, so all sources were
included. Of the 402 sources in the complete sample, 294 were selected
for VSOP observations, and this sample is designated as the VSOP
Source Sample (VSS) \citep{hir00b,edw02}.
Observations of the VSS were made between August 1997 and October 2003.
Of the VSS sample all but 29 were observed.  Fig.~1 
graphically presents the outcomes; not observed, failed,
no space fringes or sucessful.

The VSOP survey observations were made at 5 GHz, with two
left-circularly polarized 16\,MHz IF bandwidths, sampled with two bits
\citep{hir00a}. 
GOT observations of survey sources which were made with a similar
configuration, were also included (see P-III for discussion of this).
Data were usually correlated at either the DRAO Penticton correlator
\citep{car99} (54\%) or the NAOJ Mitaka correlator \citep{shi98}
(18\%), with one non-GOT experiment processed at the Socorro
correlator \citep{nap94} along with two dozen GOT extractions and a
test experiment (a total of 28\%) \citep[see][for
details]{dod08}. After correlation, the data were sent to ISAS for
distribution to the Survey Reduction Team members.

\section{Data Reduction}

The data were imported
into AIPS \citep{aips}, amplitude calibrated (with the measured or expected system
temperature and, if needed, the autocorrelation normalised) then
fringe fitted. 
After satisfactory delay and rate calibration, the data for all
spectral channels were summed to a single channel per 16~MHz sub-band
(i.e. two) and exported to DIFMAP \citep{she97} for self calibration
and model fitting. Scripts were used as much as possible to ensure
that the methods were standardized.

For the entire VSOP survey programme, 265 of the 294 sources were
observed. The observations are listed in P-III Table~1 and P-V
Table~2, which includes source names, experiment code, Ground Radio
Telescopes, Tracking stations and Correlator used, time over which
fringes were detected and the optical ID and redshift. Table~1 in P-V
contains the entire summary of all the VSS targets, whether observed
or not, with contemporary values of total density flux at 5~GHz (where
avaliable), the redshift, relevant references, best fit (or lower
limit) observer frame brightness temperatures of the core, detected
area, and flux density on the longest baselines.  For some of the
observed sources, fringes to the spacecraft were not detected.  Many
of the sources were significantly resolved on shorter ground-only
baselines, so that the lack of space fringes (RMS detection is
typically 0.1 Jy) is consistent with the resolving structure seen on
shorter baselines.  However, for others, ground observations suggested
that the space baselines (typically greater than 150 M$\lambda$)
should have been detected.  These were considered failures, with the
reasons unknown.

\section{The Results}
\subsection{The \uv\ Coverage, Visibility Amplitudes and Images}

The graphical results for most of the survey sources
are given in P-III Fig.~2 and P-V Fig.~1, 
which shows the \uv\ coverage, the visibility amplitude versus
projected \uv\ distance, and the image displayed in contour form.  
The VSOP data were able to indicate the strength and angular size of a
core component, even if, in some cases, most of the most extended
emission, shown with lower-resolution images, was resolved out in the
VSOP data.  

\citet{lis01} investigated the effect of the limited \uv\ sampling on
the imaging for HALCA and the VLBA, with simulations. In \citet{lis00}
the effects on fidelity of using few ground baselines with HALCA
(i.e. Survey observations) was investigated by comparison of survey
datasets with the complete GOT dataset from which it was drawn. The
conclusions therein were that: due to poor CLEAN deconvolution
stability image fidelity was about 30:1 to 100:1, that the Survey
datasets would have poorer dynamic range, yet would give reasonable
measures of the core brightness temperatures. Hence we expect a
typical image fidelity of 20:1. 

\subsection{The Brightness Temperature distribution}

A histogram depicting the brightness temperature distribution in
the source frame for the sources with known redshifts is
shown in Fig.~2. 
Most cores have $T_b>10^{11}$~K, with approximately 56\% of the
sources having a measured brightness temperature in excess of
$10^{12}$~K in the source frame and 8\% have greater than $5 \times 10^{12}$~K.
The
distribution presented in \citet{kov05} is from VLBA observations at
15~GHz.
They find also a median value of $10^{12}$K, but the distribution
towards $10^{13}$ and beyond is largely made up of lower limits,
rather than actual measurements as we have here. We have compared the
$T_b$ for the source in common with \citet{kov05} by selecting the
data with the closest observation dates.  The $T_b$ in the VSS tend to
be higher, as expected since the majority of the brightness
temperatures in \citet{kov05} are lower limits (70\% of the compared
sources), with a median ratio of 2.4.  Detailed comparison of
individual sources, in particular those with very different $T_b$,
will be presented in a future paper.

\subsection{The Cumulative Visibility}

We have repeated the analysis of \citet{hor04} on the complete VSS
dataset, but have yet to fully understand the differences we find. We
formed the cumulative visibilities from the scalar average of
the amplitudes, binned by \uv\ radius. We include the bias correction,
due to the scalar averaging. We will continue to work on this. 

\subsection{The Brightness Temperatures vs IDV}

IDV and T$_{\rm b}$ are both measures of compactness, as is the
detected core area, A. We were given early access to the MASIV dataset
\citep{masiv}, which we searched for correlations between their IDV
measure ($\mu$) and our measures; A and T$_{\rm b}$. Unfortunately the
errors on these values swamps any detected correlation. We will
explore the use of more robust statistical analyses to attempt to
tease out the correlations.

\section{VSOP Survey II}

For scheduling reasons we can expect that for a considerable fraction
of the mission lifetime there will be very limited ground radio
telescope avaliability. It is important to use the satellite time as
profitably as possibility, therefore we consider what might be the
best use of such limited baselines. Two major problems affected the
VSS experiments; limited GRT coverage and poor amplitude
calibration. Furthermore a repeat of a survey of AGN cores with Space
baselines at a different frequency does not have a high science
return. The maximum Brightness Temperature measurable depends only on
the physical baseline length, which is the same for VSOP-2 as it was
for HALCA. Therefore we propose the most suitable target for a survey
would be \water\ masers. The autocorrelation can be used for amplitude
calibration and the maser structures in individual channels are
usually simple, which fits the constraints for the likely Survey-II
configuration. The obvious science target for such a minimal array
would be to make finely sampled observations of the proper motion of the
maser components.

\section{Conclusions}

The imaging portion of the VSOP Survey has been completed with 242 or
the 294 sources modelfitted to measure the brightness temperature of
the core. The distribution of these T$_{\rm b}$ confirms the
distributions previously found: about 50\% greater than $10^{12}$, but
less than 10\% are greater than $5\times 10^{12}$. 



\begin{figure}
\dfig{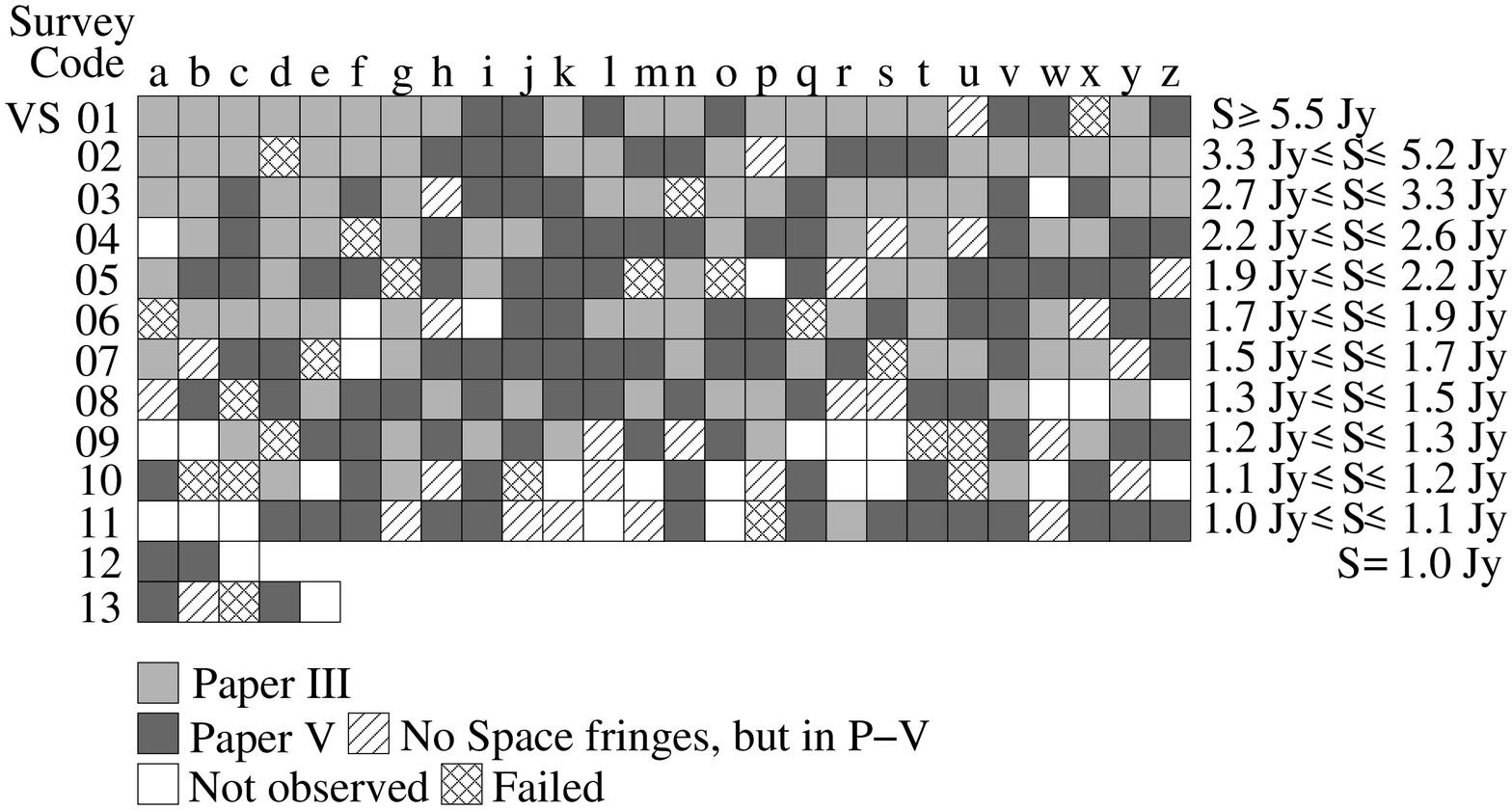}{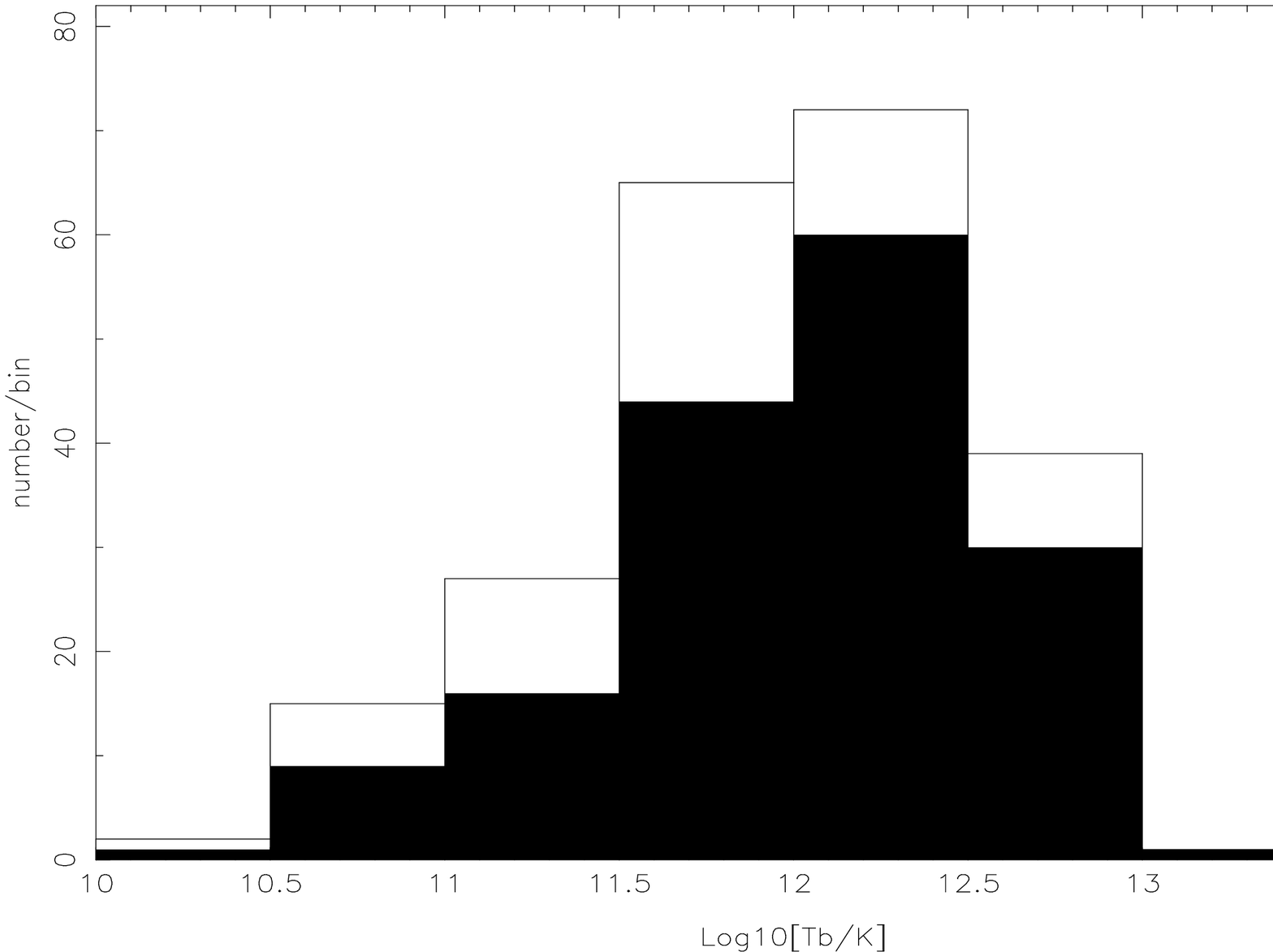}

{Fig.~1) Graphical representation of the VSS, by source code, and the
  final status: reported in which paper, whether space fringes were
  detected, or whether observations failed or were not made.\\
Fig.~2) The distribution of $T_b$ in the source frame for
the subset of 222 sources which also have a measured
redshift. Measurements are shown with a filled bar, whilst lower
limits are shown with an open bar.}
\end{figure}

\begin{footnotesize}

\end{footnotesize}

\end{document}